\documentclass[journal,final , onecolumn]{IEEEtran}

\normalsize

\usepackage{cite}
\usepackage{xcolor}
\usepackage{tikz}
\usepackage{subfigure}
\usepackage{graphicx}
\usepackage{float}
\usetikzlibrary{arrows,shapes,chains}
\usepackage{amssymb}
\usepackage{mathrsfs}
\usepackage[cmex10]{amsmath}
\usepackage{algorithm}
\usepackage{algorithmic}
\usepackage{array}
\usepackage{booktabs}
\usepackage{diagbox}
\usepackage{caption}
\begin{document}
\title{Computing the Rate-Distortion Function of Gray-Wyner System
}

\author{Guojun~Chen,~\IEEEmembership{Student Member,~IEEE,}~Yinfei~Xu,~\IEEEmembership{Member,~IEEE,}
Ling~Liu,~\IEEEmembership{Member,~IEEE,}~Tiecheng~Song,~\IEEEmembership{Member,~IEEE,}
and~Jing~Hu,~\IEEEmembership{Member,~IEEE}
\thanks{Guojun~Chen,~Tiecheng~Song and Jing~Hu are with the School of Information Science and Engineering, Southeast University, Nanjing, 210096, China, and also with the National Mobile Communication Research Laboratory, Southeast University, Nanjing 210096, China (e-mail: \{guojunchen; songtc; louy\}@seu.edu.cn).
Yinfei~Xu is with the School of Information Science and Engineering, Southeast University, Nanjing, 210096, China (e-mail: yinfeixu@seu.edu.cn).
Ling Liu is with the College of Computer Science and Software Engineering, Shenzhen University, Shenzhen 518060, China (e-mail: liulingcs@szu.edu.cn).
}
}


\maketitle

\begin{abstract}
  In this paper, the rate-distortion theory of the Gray-Wyner lossy source coding system is investigated. For the case of jointly Gaussian distributed sources, we establish an expression for the rate-distortion function under the constraint of quadratic distortion. Using the proposed rate-distortion function, any corner point on the rate-distortion region can be conveniently calculated. We take Wyner's common information as an example and provide a general and simple method to solve this problem. Through the analysis of the rate-distortion function, the rate on each layer and covariance matrix of auxiliary random variables and the sources are also presented in this paper. 
\end{abstract}


\IEEEpeerreviewmaketitle

\section{Introduction}\label{secI}
Consider that lossy source coding is a very important technique in contemporary communications to provide a greater transmission rate. The rate-distortion (RD) region of lossy source coding remains elusive after decades of research. Gray and Wyner proposed the Gray-Wyner lossy source coding system in \cite{gray1974source}, shown in Fig. \ref{fig:Gray}. However, the RD region was calculated and determined only for the case of some simple discrete sources with equal distortion constraints.

Most of the research on Gray-Wyner lossy source coding systems focused on Common Information (CI), which is a special corner point in the RD region. The two most popular CI are Wyner's CI and G{\'{a}}sc-K{\"{o}}rner CI, presented in \cite{Wyner1975common} and \cite{G1973Common}, respectively. Xu \textit{et al.} provided a converse proof of Wyner's CI, which is equal to the minimum common message rate when the total rate is arbitrarily close to the RD function under suitable conditions in \cite{xu2015lossy}. Moreover, Viswanatha \textit{et al.} was more interested in the achievability part of CI and studied both Wyner's CI and G{\' a}sc-K{\"o}rner CI in \cite{Viswanatha2014lossy}. \cite{sula2022} solved the rate region by constraining the sum of the two private rates, and calculated the Wyner's CI based on the proposed method and previous work on Wyner's CI in the Gray-Wyner system \cite{Gasptar2020,Sulaarxiv}. However, due to the constrains on private rates, the solution was just part of rate region of Gray-Wyner system. Our results are more general and able to calculate the whole achievable rate region. Furthermore, we solve the expression for Wyner's CI of two correlated Gaussian sources, which differs from the method applied in \cite{Viswanatha2014lossy}. Some partial results were studied including Wyner's CI characterization for bivariate Gaussian sources in \cite{SC}, and total correlation characterization for vector Gaussian sources in \cite{VG}.

Besides the Gray-Wyner lossy source coding system, other multi-terminal RD problems are studied in different distributed scenarios. In \cite{xiao2005compression}, Xiao \textit{et al.} considered the problem of centralized/distributed data compression via one/two encoders and only one decoder with bivariate correlated Gaussian sources. In \cite{yang2011rate,yang2014information}, Yang \textit{et al.}  studied a multi-terminal sequential data compression system whose main motivation is the causal video the coding strategy in practical standards. Furthermore, as a special case of the Gray-Wyner source coding problem, the achievability of the RD region for successive refinement problem is solved \cite{nayak2009successive}. Tuncel and Rose provided an iterative algorithm to compute the minimum sum rate and developed Kuhun-Tucker optimal conditions for successive refinement problems in \cite{tuncel2003computation}. Afterwards, Nayak and Tuncel obtained the analytical representation of the RD function, for bivariate correlated Gaussian sources in \cite{nayak2009successive}, which inspirits us in this paper. The vector form was further studied in \cite{nayak2010successive}.

Despite extensive research on lossy source coding problems, the analytical determination on the Gray-Wyner lossy source coding system is lacking, even for the simple bivariate correlated Gaussian sources. Inspirited by \cite{tuncel2003computation,nayak2009successive}, we establish an explicit formulation of the RD function for the Gaussian Gray-Wyner system with quadratic individual distortion constraints. The main contributions of this paper are summarized as follows:
\begin{itemize}
    \item We establish the achievability part for the lossy Gray-Wyner rate region with Gaussian sources. The analytical expression of the RD function for the Gaussian Gray-Wyner System under the constraint of quadratic distortion is presented in this paper. With the proposed RD function, the rate region of the lossy Gray-Wyner system is conveniently depicted.
    \item Utilizing the proposed RD function, the analytical expression of all corner points on the rate region is convenient to calculate. As an example, we recompute the known Wyner's CI with an alternative method to confirm the effectiveness of the proposed technique.
\end{itemize}

The rest of this paper is organized as follows. The system description and problem formulation are given in Section II. In Section III, we present the main results and main contributions. And the proofs of the two theorems are established in Section IV and V, respectively. 
At last, we make a summary of this paper in Section VI.


\section{System Model and Problem Formulation}
\subsection{System Model of Lossy Gray-Wyner Network}

\begin{figure}
  \centering
  \scriptsize
  \tikzstyle{format}=[rectangle,draw,thin,fill=white]
  \tikzstyle{point}=[coordinate,on grid,]
  \resizebox{0.6\textwidth}{!}{
  \begin{tikzpicture}[scale=100]
  \node[format] (Encoder1){Encoder 1};
  \node[format , below of = Encoder1 , node distance = 7mm ] (Encoder0){Encoder 0};
  \node[format , below of = Encoder0 , node distance = 7mm ] (Encoder2){Encoder 2};
  \node[format , right of = Encoder1 , node distance = 20mm ] (Decoder1){Decoder 1};
  \node[format , right of = Encoder2 , node distance = 20mm ] (Decoder2){Decoder 2};
  \node[point , right of = Encoder0 , node distance = 20mm ] (point1)  {};
  \node[point , left of  = Encoder0 , node distance = 10mm] (pointe0) {};
  \node[point , left of  = Encoder1 , node distance = 10mm] (pointe1) {};
  \node[point , left of  = Encoder2 , node distance = 10mm] (pointe2) {};
  \node[point , left of  = pointe0 , node distance = 10mm] (pointxy) {};
  \node[point , right of  = Decoder1 , node distance = 12mm] (pointhatx) {};
  \node[point , right of  = Decoder2 , node distance = 12mm] (pointhaty) {};
  \draw[ -> ](Encoder1)--node[above] {$V_1$} (Decoder1);
  \draw[ -> ](Encoder2)--node[above] {$V_2$} (Decoder2);
  \draw[ -  ](Encoder0)--node[above]{$U\qquad\;$}(point1);
  \draw[ -> ](point1)--(Decoder1);
  \draw[ -> ](point1)--(Decoder2);
  \draw[ -> ](pointe0)--(Encoder0);
  \draw[ -> ](pointe1)--(Encoder1);
  \draw[ -> ](pointe2)--(Encoder2);
  \draw[ -  ](pointxy)--node[above]{$X_1,X_2$}(pointe0);
  \draw[ -  ](pointe0)--(pointe1);
  \draw[ -  ](pointe0)--(pointe2);
  \draw[ -> ](Decoder1) -- node [right] {\quad $(\hat X_1, D_1)$} (pointhatx);
  \draw[ -> ](Decoder2) -- node [right] {\quad $(\hat X_2,D_2)$} (pointhaty);
  \end{tikzpicture}}
  \caption{Gray-Wyner lossy source coding system.}
  \label{fig:Gray}
\end{figure}
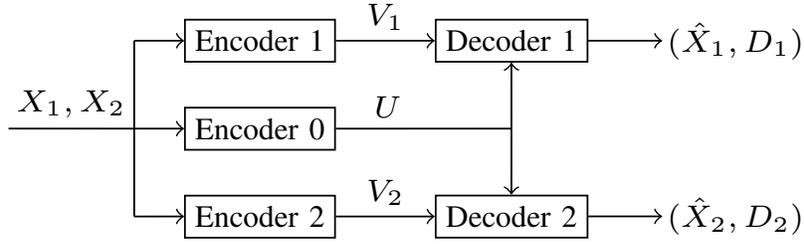

We consider a point-to-point communication system as depicted in Fig. \ref{fig:Gray}. In this system, memoryless correlated sources are written by the sequence $\{(X_1,X_2)^n\}_{n=1}^{\infty}$, where each pair of random variable takes value in a finite state alphabet ${ {\cal X}}_1 \times {\cal X}_2$ with joint probability mass function $p(x_1,x_2)$. We wish to communicate over the Gray-Wyner network to transmit the source message and reconstruct the codeword with distortion constraints. A $(2^{nR_i}, n)$ code, where the rate of this code is the real number $R_i>0$, $i =  0,\ 1,\ 2$, consists of
\begin{itemize}
 \item Three encoders have access to both sources and produce three message sets $([1:2^{nR_i}])$ during encoding functions $f_i^{(n)}$:
 \begin{align*}
  f_i^{(n)} : {\cal X}_1^{(n)} \times {\cal X}_2^{(n)}\rightarrow \{  1,2, \cdots ,  2^{nR_i}  \}, \quad i =  0,\ 1,\ 2 .
 \end{align*}
\item  Each of two decoders is only interested in reconstructing one source and has access to only two messages, where Decoder $i$ has access to messages $M_0 \in [1:2^{nR_0}]$ and $M_i \in [1:2^{nR_i}]$, with decoding function $g_i^{(n)}$:
\begin{align*}
  g_i^{(n)} :  M_0 \times M_i \rightarrow \hat { {\cal X}}_i^{(n)},  \quad i = 1, \ 2 .
 \end{align*}
 where the reconstructions, denoted as $\{ \hat X_i \}_{n=1}^{\infty}$, are sequences of symbols in alphabets $\hat { {\cal X}}_i$.
 \item Decoder $i$ is only interested in reproducing $\hat X_i$ with constraint average distortion $D_i$, considering the following mapping as the distortion measure
 \begin{align*}
  d_i : {\cal{X}}^{(n)}_i \times \hat{{\cal{X}}}^{(n)}_i  \rightarrow [ 0 , \infty ) , \quad  i = 1, \ 2 .
\end{align*}
\end{itemize}

  A quintuple tuple $(R_0 , R_1 , R_2 , D_1 , D_2)$ is said to be achievable if there exists a sequence of encoding functions $f_i^{(n)}$ and decoding functions $g_j^{(n)}$, satisfy
 {\setlength\abovedisplayskip{1pt}
\setlength\belowdisplayskip{1pt}
  \begin{align*}
    \limsup_{n \rightarrow \infty} & \frac{1}{n} \log |{\cal M}_i^{(n)}|  \leq R_i, \quad i = 0 ,\ 1, \ 2  ; \\
    \limsup_{n \rightarrow \infty} &   \mathbb{E} \left[ d_j (X_j^{(n)} , g_j^{(n)}(M_{0}^{(n)}, M_{j}^{(n)}   )  )  \right]  \leq D_j, \quad j = 1, \ 2 .
  \end{align*}}
Next, we will calculate the RD region under this system model.

\subsection{Problem Expression of Rate-Distortion Region}\label{II.B}
In this section, we first give an alternative single-letter characterization of the rate region and then define the weighted sum rate as the RD function. To solve the minimum value of the RD function, a multivariate optimization problem with some auxiliary variables is considered. In subsequent sections, we will focus on solving this optimization problem in detail.

The RD region $\mathscr{R}^*$ for the Gray-Wyner system is the closure of the set of achievable rate triples referred \cite[Th. 14.3]{el2011network}. For ease of understanding, we present the single-letter characterization of the lossy Gray-Wyner system rate region in the following lemma.

\newtheorem{lemma}{Lemma}
\begin{lemma}[Single-letter Characterization]\label{singalletter}
  A tuple $(R_0 , R_1 , R_2 )$ is said to be achievable subject to distortion constraints $(D_1,D_2)$ if and only if there exist auxiliary random variables $U$ and $(\hat X_1,\hat X_2)$ jointly distributed with $(X_1 , X_2)$ satisfying
\begin{align}
  \label{orignalcondition_1}& R_0 \geq I ( X_1 , X_2 ; U ),             \\
  \label{orignalcondition} & R_i \geq I (X_1, X_2 ; \hat X_i | U),  \qquad   i = 1, \ 2     \\
  \label{orignalcondition_2}& D_i \geq  \mathbb{E} [ d_i ( X_i , \hat X_i ) ].
\end{align}
\end{lemma}
\begin{IEEEproof}
According to \cite[Th. 7]{gray1974source}, \cite[Sec. II. A]{nayak2009successive}, \cite[Ch. 14.2]{el2011network} and \cite[Eq. (5)]{Viswanatha2014lossy}, the original conditions of \eqref{orignalcondition} that RD tuples $(R_0 , R_1 , R_2 ,D_1,D_2)$ satisfied are $ R_i \geq I (X_i; \hat X_i | U), \,  i = 1, 2  $. Because the reproduction $\hat X_i$ is the function of source $X_i$ and auxiliary variable $U$, this lemma is easy to be proved using the fact $I (X_2; \hat X_1 | U, X_1) = I (X_1; \hat X_2 | U, X_2) = 0$.
\end{IEEEproof}

In a similar spirit as the Arimoto-Blahut Algorithm in \cite{blahut1972computation,secondorder1}, it is convenient to consider a multivariate optimization problem with some auxiliary variables to calculate the RD region. Therefore, the above single-letter characterization is transmitted into a weighted sum function, which is the entire tangent of the rate region.
Probability density function $p_V(v)$ is replaced by $p(v)$. To make the paper simplified, we first denote the vector symbol $\boldsymbol{\alpha} = (\alpha_0,\alpha_1,\alpha_2)$ with $\alpha_i \geq 0 $ for $i =0,1,2$ and $\boldsymbol{\beta} = (\beta_1,\beta_2)$ with $\beta_i \geq 0 $ for $i =1,2$. Then, we define this weighted sum rate under individual distortion constraint as RD function $ R_{\boldsymbol{\alpha}} (D_1,D_2) $ via
\begin{align}\label{RDfunction}
    R_{\boldsymbol{\alpha}} (D_1,D_2) \triangleq  \mathop {\min }\limits_{ \scriptsize {\begin{array}{ccc}
      p({u, \hat x_1, \hat x_2| x_1 , x_2 }): \\
      \mathbb{E} [ d_1 ( X_1 , \hat X_1 ) ] \leq D_1 \\
      \mathbb{E} [ d_2 ( X_2 , \hat X_2 ) ] \leq D_2
      \end{array}} } \alpha_0 I(X_1 , X_2 ; U)  + \sum_{i=1}^{2} \alpha_i I (X_1, X_2 ; \hat X_i | U).
  \end{align}
With the help of the RD function, the RD region can be obtained simply by solving the above optimization problem. However, even though the RD region is transferred to a function with mutual information, the computation is still a major and indirect problem.
  Since the convexity of this region is proved referred to \cite{tuncel2003computation} and \cite{boyd2004convex}, the Lagrangian function $\mathcal{L}^P$, which is the function of $p({\hat x_1, \hat x_2, u| x_1 , x_2 })$, is applied
\begin{align}\label{Lagrangian_function}
    \mathcal{L}^P  \triangleq \alpha_0 I(X_1 , X_2 ; U)
     + \sum_{i=1}^{2} \alpha_i I (X_1, X_2 ; \hat X_i | U)  + \beta_i  \mathbb{E} [ d_i ( X_i , \hat X_i ) ].
  \end{align}
And  we define the minimum value of this Lagrangian function as $\mathcal{L}^*$ with the constraint of $\mathbb{E} [ d_1 ( X_1 , \hat X_1 ) ] \leq D_1 $ and $      \mathbb{E} [ d_2 ( X_2 , \hat X_2 ) ] \leq D_2$.

Therefore, referred to \cite{secondorder2}\cite{secondorder3}, the closure of all sets of $(D1, D2)$-achievable rate triplets is
  \begin{align}\label{RDmax}
    R_{\boldsymbol{\alpha}} (D_1,D_2) =  \mathop {\max }\limits_{{\beta _1},{\beta _2}} \{ \mathcal{L}^* - {\beta _1}{D_1} - {\beta _2}{D_2}  \}.
  \end{align}

In the rest of this paper, we will focus on computing the optimal RD function $R_{\boldsymbol{\alpha}} (D_1,D_2)$.

\section{Main Results and Analytical Expression of RD Function}\label{secIII}
In this section, we first propose some solutions to the RD function for degenerate conditions, which have been solved previously. Then, focusing on general conditions, an important and interesting result is raised in Theorem 1. Furthermore, the analytical expression of the RD function is put forward by Theorem 2 with a piecewise function. Finally, we recompute the Wyner's CI through the proposed RD function, which is different from that in \cite{Viswanatha2014lossy}.

Gaussian sources with linear Gaussian encoders and decoders are wildly used in source coding problems and squared-error distortion measurement is one of the most popular measure methods in practice. In this paper, we focus on calculating the RD region for Gaussian conditions, which is the achievability part of the RD region for the lossy Gray-Wyner lossy coding system. Under the assumption of Gaussian encoders and decoders, the auxiliary random variable $U$ is also Gaussian distributed.
For completeness, the RD function for simple special cases is raised in the following lemma. And in subsequent sections, we will focus on the non-trivial condition, denoted as ${\cal D}_{\boldsymbol{\alpha},\boldsymbol{D}} ^{NZ} \triangleq \{ ( \boldsymbol{\alpha} , D_1 , D_2 ) :\boldsymbol{\alpha} > 0 ,  \alpha_1  < \alpha_0, \  \alpha_2 < \alpha_0 ,\ \alpha_1 +\alpha_2  > \alpha_0 , D_1 < 1, D_2 < 1 \} $.
\newtheorem{remark}{Remark}
\begin{remark}\label{remark1}
For ease of understanding, the value of $ \alpha_i$ is considered to be the communication cost coefficient on each communication channel, and the RD function is the summed communication cost function. Therefore, in order to minimize the sum communication cost, when $\alpha_1  > \alpha_0$, no messages are transmitted through the first private channel, and the problem degenerates into a successive refinement problem, which was studied and discussed in \cite{gray1973,equitz1991successive}. Furthermore, when $\alpha_1 +\alpha_2  < \alpha_0$, no messages are transmitted through the common channel, and the problem degenerates into two separate point-to-point communication problems.
\end{remark}

\begin{lemma}\label{lemma:simplecase}
 The analytical expression of RD function for Gaussian Gray-Wyner lossy source coding system in some degenerated cases is as follows
\begin{small}
\begin{align}\label{Reasy}
  \begin{array}{*{20}{l}}
  R_{\boldsymbol{\alpha},SPECIAL}(D_1,D_2) =  \left\{
    {\begin{array}{*{20}{l}}
     0
    & if  \  \alpha_0 = 0  \\
    \frac{   \alpha_0   } {2} \log \frac {1} { D_1 }
    &if   \  \alpha_2 = 0, \alpha_1 > \alpha_0 > 0 \ or \ D_2 \ge 1            \\
    \frac{  \alpha_0   } {2} \log \frac {1} { D_2 }
    &if  \ \alpha_1 = 0, \alpha_2 > \alpha_0 > 0 \ or \ D_1 \ge 1            \\
    \frac{  \alpha_1   } {2} \log \frac {1} { D_1 } + \frac{  \alpha_2   } {2} \log \frac {1} { D_2 }
    &if   \ \alpha_1 + \alpha_2 \le \alpha_0 \\
    \frac{1-\rho^2}{ D_1 D_2 - (\rho - \sqrt{(1-D_1)(1-D_2)})^2 }
    & if \ \alpha_0 < \alpha_1, \alpha_0 < \alpha_2 \\
    R_{SR}(D_1,D_2)
    &if \ \alpha_1 < \alpha_0 < \alpha_2 \ or \ \alpha_2 < \alpha_0 < \alpha_1 \\
  \end{array}} \right.
  \end{array}
\end{align}
\end{small}
where $R_{SR}(D_1,D_2)$ is the RD function for successive refinement, solved in \cite[Theorem 2]{nayak2009successive}.

\end{lemma}
\begin{IEEEproof}
The first four terms are obtained directly from the simplest point-to-point lossy source coding problem and RD function. And for the fifth term, the problem of the Gray-Wyner system degenerates into the problem of computing the RD function of correlated Gaussian sources under the individual distortion criterion, which has been solved in \cite[Theorem 6]{xiao2005compression}. At last, when $(\alpha_0, \alpha_1, \alpha_2)$ satisfies the sixth condition, the problem degenerates into successive refinement, which has been solved in \cite[Theorem 2]{nayak2009successive}.
\end{IEEEproof}

\subsection{ Intermediate Results of RD Function}
In this section, we transform the original optimization problem \eqref{RDmax} into a simple function with only three parameters, namely two auxiliary variables $m_1$, $m_2$ and the variance of random variable $U$.  For simplicity, we first define two binary cubic functions via
  \begin{align}
    \label{f1_the2}& f_{\alpha_1} (m_1, m_2)  =  \frac{\alpha_1}{\alpha_0} (m_2 -m_1 \rho ) (\rho - m_1 m_2 \sigma^2_U) + (\frac{\alpha_1}{\alpha_0} -1) (m_1 -m_2 \rho) (1- m_1^2 \sigma^2_U) , \\
   \label{f2_the2}& f_{\alpha_2} (m_1, m_2)  =  \frac{\alpha_2}{\alpha_0} (m_1 -m_2 \rho ) (\rho - m_1 m_2 \sigma^2_U) + (\frac{\alpha_2}{\alpha_0} -1) (m_2 -m_1 \rho) (1- m_2^2 \sigma^2_U)   ,
  \end{align}
where $\rho$ is the correlation coefficient of two sources $X_1$ and $X_2$, and $ \sigma^2_U$ is the variance of the random variable $U$. Then we begin to exhibit the intermediate results of the RD function.

\newtheorem{theorem}{Theorem}

\begin{theorem}\label{simplyR}
 For arbitrary parameters $\boldsymbol{\alpha}$ and distortion constraint $(D_1, \, D_2)$ satisfying $(\boldsymbol{\alpha}, \, D_1, \, D_2) \in {\cal D}_{\boldsymbol{\alpha},\boldsymbol{D}} ^{NZ}$, the RD function is a function of auxiliary variables $m_1$, $m_2$ and $U$ via:
 \begin{small}
  \begin{align}
    \label{Rcom}
    &R_{\boldsymbol{\alpha},GENERAL}(D_1,D_2)
     = \frac{\alpha_0}{2} \log \frac{1-\rho^2}{(1- m_1^2 \sigma^2_U) (1 - m_2^2 \sigma^2_U) - (\rho - m_1 m_2 \sigma^2_U)^2}
    + \sum_{i=1}^2 \frac{\alpha_i}{2} \log \frac{1-m_i^2\sigma^2_U }{D_i}, 
  \end{align}
  \end{small}
where the parameters $m_1$, $m_2$ and $\sigma^2_U$ satisfy
\begin{align}
  \label{gamma1_the2}& f_{\alpha_1} (m_1, m_2) ( 1- m_1^2 \sigma_U^2 - D_1 ) = 0, \\
  \label{gamma2_the2}& f_{\alpha_2} (m_1, m_2) ( 1- m_2^2 \sigma_U^2 - D_2 ) = 0, \\
  \label{con2}&  m_1 m_2 \sigma^2_U \leq \rho < \min\{ \frac{m_1}{m_2} , \frac{m_2}{m_1} \}  ,
\end{align}

\begin{IEEEproof}
   Detailed proofs are provided in Section \ref{Proof_th1}. Here, we only give a brief process of the complete proof. First, we transform the optimization problem $\mathcal{L}^{*}$ into a dual optimal form and establish an alternative characterization for $\mathcal{L}^{*}$ by solving the dual optimal problem. Then, we focus on the characterization of Gaussian random variables and set up a system of equations to compute the values of all parameters in $\mathcal{L}^{*}$. At last, the RD function is obtained when variable substitution is performed with \eqref{RDmax}.
   \end{IEEEproof}
\end{theorem}

The importance of this intermediate result is that we derive the optimization problem of the RD function in \eqref{RDmax} to this computable formula, and we only need to discuss three parameters $m_1$, $m_2$ and $\sigma^2_U$ to solve the RD function. Moreover, it is the general formula of the RD function for the Gray-Wyner system. With this form, we discuss the rate on each channel and the covariance matrix for the sources $X_1, \, X_2$ and auxiliary random variable $U$ in the remarks below.

\begin{remark}
The solution of the RD function is meaningful to derive the rate on each communication channel. According to \eqref{Rcom}, the RD region is distributed referred to
\begin{small}
\begin{align}
 \label{R0} &I( X_1, X_2 ; U ) = \frac{1}{2} \log \frac{1-\rho^2}{(1- m_1^2 \sigma^2_U) (1 - m_2^2 \sigma^2_U) - (\rho - m_1 m_2 \sigma^2_U)^2},\\
 \label{R1}&I( X_1, X_2; \hat X_1 | U ) = I( X_1; \hat X_1 | U ) = \frac{1}{2} \log \frac{1-m_1^2\sigma^2_U }{D_1}, \\
 \label{R2}&I( X_2, X_2; \hat X_2 | U ) = I( X_2; \hat X_1 | U ) = \frac{1}{2} \log \frac{1-m_2^2 \sigma^2_U }{D_2}.
\end{align}
\end{small}
Moreover, we can obtain the optimal $p(x_1,x_2,u)$, which is the Gaussian probability density function with covariance matrix $\mathrm{Cov}(X_1,X_2,U)$ via
\begin{align}\label{xxu}
\mathrm{Cov}(X_1,X_2,U) = \left[ {\begin{array}{*{10}{c}}
     1   &  \rho  &   m_1 \sigma^2_U  \\
     \rho    &   1  &  m_2 \sigma^2_U \\
     m_1 \sigma^2_U   &  m_2 \sigma^2_U & \sigma^2_U \\
    \end{array}} \right].
\end{align}
From this covariance matrix, the physical interpretations of the auxiliary parameters $m_1$, $m_2$ and $\sigma_U^2$ are obvious. $m_1$ is considered as the correlation coefficient between the source $X_1$ and the auxiliary random variable $U$. For $m_2$, it is symmetric.
\end{remark}

\begin{remark}
Compared to successive refinement, it is a degenerated form of our Gray-Wyner system, because the Gray-Wyner system degenerates to successive refinement when one of the private channels is removed. From a formulation perspective, as shown in the last line of \eqref{Reasy}, the Gray-Wyner system degenerates to successive refinement when letting $\alpha_1 < \alpha_0 < \alpha_2$ or $\alpha_2 < \alpha_0 < \alpha_1 $. Meanwhile, this theorem can also confirm that the solution of the RD function is the same as successive refinement, as shown in \cite[Eq.(71)]{nayak2009successive}, under specific situations.
\end{remark}

\begin{remark}\label{Remark4}
The values of reconstruction source $\hat X_1$ and $\hat X_2$ according to the proof of Theorem 1 in Section \ref{Proof_th1}. Intuitively, the private channel encoded message $V_1$ only has access to reconstruct $\hat X_1$ and is a zero mean Gaussian random variable. Because the scale of Gaussian random variable has no effect, it is valid to set $\sigma_{V_1}^2 =  \sigma_{X_1}^2$. For $\sigma_{V_2}^2$, it is symmetric. Therefore, with the solution of $m_1$, $m_2$ and $\sigma_U^2$, the reconstruction sources $\hat X_1$ and $\hat X_2$ are established, and RD function is solved analytically.
\end{remark}

\subsection{Analytical Expression of RD Function}
From the RD function in \eqref{Rcom}, we need to determine the value of the parameters set $( m_1 , m_2 , \sigma^2_U  )$ for different distortion pairs $(D_1, D_2)$ and parameters $ (\alpha_0, \alpha_1, \alpha_2) $ when solving the RD Function. According to the expression of \eqref{Reasy} and \eqref{Rcom}, the analytical formulation of the RD function is proposed in the following theorem, and the proof is shown in subsequent sections.

\begin{theorem}\label{finalRRR} Analytical expression of the RD function for Gaussian Gray-Wyner lossy source coding system is
  \begin{align}
    R_{\boldsymbol{\alpha}}(D_1,D_2) =
    \left\{ {\begin{array}{*{20}{l}}
    R_{\boldsymbol{\alpha},GENERAL}(D_1,D_2)
    & if  \  (\boldsymbol{\alpha},\boldsymbol{D}) \in {\cal D}_{\boldsymbol{\alpha},\boldsymbol{D}} ^{NZ}\\
    R_{\boldsymbol{\alpha},SPECIAL}(D_1,D_2)
    & if  \  (\boldsymbol{\alpha},\boldsymbol{D})  \notin {\cal D}_{\boldsymbol{\alpha},\boldsymbol{D}} ^{NZ}
    \end{array}} \right. ,
  \end{align}
  where
  \begin{small}
    \begin{align}
    (m_1, m_2 , \sigma_U^2) =
    \left\{ {\begin{array}{*{20}{l}}
      (1, m_2^* , 1-D_1  )
      & if  \  (D_1,D_2) \in D_{\boldsymbol{D}}^{1}  \\
      ( m_1^* , 1, 1-D_2 )
      & if \ (D_1,D_2) \in D_{ \boldsymbol{D}}^{2}  \\
      ( \sqrt{1-D_1} , \sqrt{1-D_2} , 1  )
      &if   \ ( D_1,D_2) \in D_{ \boldsymbol{D}}^{3}             \\
      ( \sqrt{1-\nu_1} , \sqrt{1-\nu_2} , 1  )
      &if   \  (D_1,D_2) \in D_{ \boldsymbol{D}}^4 ,   \alpha_1 \neq \alpha_2          \\
      (\sqrt{ \left| \frac {\alpha_1 \rho + \alpha_1 - \alpha_0} {\alpha_0 - \alpha_1 - \alpha_2 } \right| } , \sqrt{ \left| \frac {\alpha_2 \rho + \alpha_2 - \alpha_0} {\alpha_0 - \alpha_1 - \alpha_2 } \right| } , 1)
      &if   \  (D_1,D_2) \in D_{ \boldsymbol{D}}^4 ,    \alpha_1 = \alpha_2        \\
    \end{array}} \right. ,
    \end{align}
    \end{small}
    and where
    \begin{small}
\begin{align}
\label{region_D1}&  D_{ \boldsymbol{D}}^1 \triangleq \left\{  (D_1,D_2) : \ D_1 > \frac { \frac{\alpha_1}{\alpha_0} ( m_{2}^* - \rho )^2 } { \frac{\alpha_1}{\alpha_0} ( m_{2}^{*2} - 2 m_{2}^* \rho + 1 ) + m_{2}^* \rho - 1 },  \ \ D_2 <  1 - ( 1 - D_1 ) m_{2}^{*2} \right\}, \\
\label{region_D2}&  D_{ \boldsymbol{D}}^2 \triangleq \left\{  (D_1,D_2) : \ D_2 > \frac { \frac{\alpha_2}{\alpha_0} ( m_{1}^* - \rho )^2 } { \frac{\alpha_2}{\alpha_0} ( m_{1}^{*2} - 2 m_{1}^* \rho + 1 ) + m_{1}^* \rho - 1 },   \ D_1 <  1 - ( 1 - D_2 ) m_{1}^{*2} \right\},  \\
\label{region_D3}&  D_{ \boldsymbol{D}}^3 \triangleq \left\{  (D_1,D_2) : f_{\alpha_1} (\sqrt{1-D_1} , \sqrt{1-D_2}) > 0, \  f_{\alpha_2}(\sqrt{1-D_1} , \sqrt{1-D_2}) > 0  \right\} , \\
\label{region_D4}& D_{ \boldsymbol{D}}^4 \triangleq \left\{  (D_1,D_2) : \ D_1 < \frac { \frac{\alpha_1}{\alpha_0} ( m_{2}^* - \rho )^2 } { \frac{\alpha_1}{\alpha_0} ( m_{2}^{*2} - 2 m_{2}^* \rho + 1 ) + m_{2}^* \rho - 1 },   \ D_2 < \frac { \frac{\alpha_2}{\alpha_0} ( m_{1}^* - \rho )^2 } { \frac{\alpha_2}{\alpha_0} ( m_{1}^{*2} - 2 m_{1}^* \rho + 1 ) + m_{1}^* \rho - 1 }  \right\} ,
\end{align}
\end{small}
with
\begin{small}
\begin{align}
\label{nu1}\nu_1 =  \frac { ( \alpha_0 - \alpha_1 ) \alpha_1  }  {( \alpha_0 - \alpha_2 ) \alpha_2 - ( \alpha_0 - \alpha_1 ) \alpha_1   } \left( \left( \frac { (\alpha_2 - \alpha_1) \rho + \sqrt { (\alpha_1 - \alpha_2)^2 \rho^2 + 4 ( \alpha_0 - \alpha_1 )( \alpha_0 - \alpha_2 )  } } {2 ( \alpha_0 - \alpha_1 ) } \right)^2  -1  \right)  ,  \\
\label{nu2}\nu_2 = \frac { ( \alpha_0 - \alpha_2 ) \alpha_2  }  {( \alpha_0 - \alpha_1 ) \alpha_1 - ( \alpha_0 - \alpha_2 ) \alpha_2   } \left(  \left(  \frac { (\alpha_1 - \alpha_2) \rho + \sqrt { (\alpha_1 - \alpha_2)^2 \rho^2 + 4 ( \alpha_0 - \alpha_1 )( \alpha_0 - \alpha_2 )  } } {2 ( \alpha_0 - \alpha_2 ) } \right)^2 - 1  \right),
\end{align}
\end{small}
 and $ m_1^* $ is the root of $f_{\alpha_1} (m_1 , \sqrt{{(1- D_2)}/{\sigma^2_U}} )$ , $ m_2^* $ is the root of $f_{\alpha_2} (\sqrt{{(1- D_1)}/{\sigma^2_U}} , m_2 )$ both satisfying \eqref{con2}.
\end{theorem}

\begin{IEEEproof}
  From the facts of Theorem \ref{simplyR}, we discuss and solve the condition in \eqref{gamma1_the2} and \eqref{gamma2_the2}, which satisfy \eqref{con2} to determine the solution of all parameters $m_1$, $m_2$ and $\sigma_U^2$. The detailed discussion and proof are provided in Section \ref{Proof_th2}.
\end{IEEEproof}

Therefore, the entire RD function of the Gray-Wyner lossy source coding system is to combine \eqref{Reasy} with the solution in Theorem \ref{finalRRR}, and the RD region is the closure form of the proposed RD function with integral choices of sets $ (\alpha_0 , \alpha_1, \alpha_2)$. 
Next, we make some remarks on this result.

\begin{remark}
To describe the RD region of the Gaussian Gray-Wyner lossy source coding system, we only need to combine the results of \eqref{R0}-\eqref{R2} with single-letter characterization \eqref{orignalcondition_1}-\eqref{orignalcondition_2} with the value of $m_1, m_2, \sigma_U^2 $ solved in this theorem.
\end{remark}
\begin{remark}
With the result in this theorem, it is convenient to obtain every corner point on the rate region. For instance, to obtain the Wyner's CI, we only need to determine $\alpha_0 = \alpha_1 = \alpha_2 = 1$.  The convenient method of calculation for Wyner's CI is proposed in detail in the next subsection. Therefore, this result provides us with a useful technique to compute the achievability of every corner point on the Gray-Wyner system and to research some new CI.
\end{remark}

\subsection{Wyner's Common Information }\label{section_common}
In this subsection, we reprove the lossy extension of Wyner's CI for bivariate Gaussian sources, which has been solved in \cite[Th.2]{Viswanatha2014lossy}. \cite{Viswanatha2014lossy} calculated the Wyner's CI by generating $(\hat X_1, \ \hat X_2)$ with $U$. The authors analyzed each regime of distortions to find the Markov chain between the sources, $U$ and reconstructed sources. At last, they derived the value of Wyner's CI by establishing an MMSE estimation and introducing auxiliary random variables with zero mean and unit variance. However, we give a new idea of computing CI, which is general and can simply obtain Wyner's CI. We consider the Wyner's CI as a corner point of the RD region, which means that we only need to find the corresponding point on the solved RD region to obtain Wyner's CI. Therefore, using our alternative method, all corner points in the RD region of the Gray-Wyner system are convenient to calculate, instead of proposing different techniques for different corner points. The purpose of this section is to demonstrate that the proposed analytical solution of the RD function is useful in applications and to verify the correctness of the proposed RD function.

According to \cite[Sec. III]{Viswanatha2014lossy}, the lossy Wyner's CI $C_{W}(D_1, D_2)$ is defined as
\begin{small}
\begin{align}
C_{W}(D_1, D_2) = \inf\{R_{0} : (R_0,R_1,R_2) \in R_{GW}(D_1,D_2),R_0 + R_1 +R_2 \leq R_{X_1, X_2} (D_1, D_2)+ \epsilon.\}.
\end{align}
\end{small}
Then this problem is transformed to the calculation of the infimum $R_0$ when satisfying RD function in \eqref{Rcom} and letting $\boldsymbol{\alpha} = (1,1,1)$. Next, we discuss the four distortion regions of $(D_1, D_2)$ as Theorem \ref{finalRRR}.
\begin{itemize}
\item Case 1: $(D_1,D_2) \in D_{ \boldsymbol{D}}^1 \cup D_{ \boldsymbol{D}}^2$ (Same as third line of \cite[(29)]{Viswanatha2014lossy}.)\\
    When $ \boldsymbol{\alpha} = (1,1,1) $ is considered, the distortion region $D_{ \boldsymbol{D}}^1$ is simplified to
    \begin{align}
    D_{ \boldsymbol{D}}^1 = \left\{  (D_1,D_2) : \frac{1-D_1}{1-D_2} < \rho^2 \right\},
    \end{align}
    with the choice of auxiliary variables set $ (m_1, m_2, \sigma_U^2) = ( 1 ,\ 1 / \rho , \ \sqrt{1- D_1}  )  $, the infimum $R_0$ is equal to RD function \eqref{Rcom}, which is
    \begin{align}
     C_{W}(D_1, D_2) = R_{\boldsymbol{\alpha}}(D_1,D_2) = \frac{1}{2} \log \frac{1} {D_2}.
    \end{align}
    symmetrically, when $(D_1,D_2) \in D_{ \boldsymbol{D}}^2$ the distortion region and infimum $R_0$ are
    \begin{align}
    & D_{ \boldsymbol{D}}^2 = \left\{  (D_1,D_2) : \frac{1-D_2}{1-D_1} < \rho^2 \right\}, \\
    & C_{W}(D_1, D_2) = \frac{1}{2} \log \frac{1} {D_1}.
    \end{align}
    \item Case 2: $(D_1,D_2) \in D_{ \boldsymbol{D}}^3$ (Same as second line of \cite[(29)]{Viswanatha2014lossy})\\
    With the condition of $ \boldsymbol{\alpha} = (1,1,1) $, the distortion region $D_{ \boldsymbol{D}}^3$ is established as
    \begin{align}
    D_{ \boldsymbol{D}}^3 = \left\{  (D_1,D_2) : {(1-D_1)}{(1-D_2)} < \rho^2 , \ \mathrm{min} \left\{ \frac{1-D_2}{1-D_1} , \ \frac{1-D_1}{1-D_2}  \right\} > \rho^2  \right\},
    \end{align}
    with the choice of auxiliary variables set $ (m_1, m_2, \sigma_U^2) = ( \sqrt{1-D_1} ,\ \sqrt{1-D_2} , \ 1  )  $. In this case, it is easy to check the value of \eqref{R0} is equal to \eqref{Rcom}. Therefore, the rate of $R_0$ is equal to the RD function via
\begin{align}
 C_{W}(D_1, D_2) = R_{\boldsymbol{\alpha}}(D_1,D_2) = \frac{\alpha_0}{2} \log \frac{1 - \rho^2} {D_1 D_2 - (\rho - \sqrt{ { (1-D_1 ) (1-D_2) } } )^2 }  .
\end{align}

\item Case 3: $(D_1,D_2) \in D_{ \boldsymbol{D}}^4$ (Same as second line of \cite[(30)]{Viswanatha2014lossy} )\\
In this case, with the value of $m_1 = m_2 \sigma^2_U = \rho$ obtained from the constraint in \eqref{p=m1m20}, the rate on common channel is
\begin{align}
  &   R_0 = \frac{1}{2} \log \frac{1-\rho^2}{(1- m_1^2 \sigma^2_U) (1 - m_2^2 \sigma^2_U)}.
 \end{align}
 Therefore, the problem of calculating infimum rate of $R_0$ transformed to calculate the supremum of function $f(x)=  1 + \rho^2 - ( x + \frac {\rho^2} {x})$ when we denote $ m_1^2 \sigma^2_U $ as $x$. It is obvious that function $f(x)$ is the hook function, and we need to discuss the value ranges of $x$ to solve the problem.

\begin{enumerate}
\item When ${\rho^2} / {(1 - D_1)} \le \rho $ and $1 - D_1 \ge \rho$, which is equal to
 $\{ (D_1,D_2):D_1 \le 1 - \rho , \  D_2 \le 1 - \rho \} $. With $f_{\mathrm{max}} ( x ) = f( \rho )$, the CI $C_{W}(D_1, D_2)$ is
 \begin{align}
  C_{W}(D_1, D_2) = \frac{1}{2} \log \frac{1+\rho}{1 - \rho }.
 \end{align}
 This situation is the same as the first line of \cite[(30)]{Viswanatha2014lossy}.
\item When $\rho <  {\rho^2} / {(1 - D_2)} < 1 - D_1$, which is equal to $\{ D_1 < 1 - \rho < D_2 , \  (1-D_1)(1-D_2) > \rho^2 \} $.  With  $f_{\mathrm{max}} ( x ) = f(  {\rho^2}/ {(1 - D_2)} )$,  the CI  $C_{W}(D_1, D_2)$ is
 \begin{align}
  C_{W}(D_1, D_2) = \frac{1}{2} \log \frac{1-\rho^2}{\rho^2 + D_2 - \frac{\rho^2}{1-D_2} }.
 \end{align}
 This situation is the same as the third line of \cite[(30)]{Viswanatha2014lossy}.
 \item Symmetric with case 2), we obtain the same result as the fourth line of \cite[(30)]{Viswanatha2014lossy}.
\end{enumerate}
\end{itemize}

 Therefore, all three cases make up the alternative method to compute lossy Wyner's CI of bivariate Gaussian sources.

\section{Proof of Theorem \ref{simplyR}}\label{Proof_th1}
In this section, we firstly transmit the optimal problem $\mathcal{L}^{*}$ into a dual optimal problem during introducing an auxiliary distribution function. Then we propose an alternative characterization for $\mathcal{L}^{*}$ by solving the dual optimal problem. Next, according to the assumption of the linear Gaussian system, we compute the connect the alternative characterization with distribution of Gaussian random variables and then provide an alternative form of covariance matrix for the source, which is equal to the given covariance matrix. Then the theorem is proved by solving the first order equations.

We first give the alternative characterization for $ \mathcal{L}^{*} $, which is the lower bound of $ \mathcal{L}^{*} $, is expressed in next lemma

\begin{lemma}\label{Lforalternative}
	There exists an alternative characterization for $ \mathcal{L}^{*} $ such that
	\begin{align}
		\label{Lfunctionoftheta}
		\mathcal{L}^{*}  =  - \alpha_0 \sum_{x_1 , x_2} p(x_1 , x_2) \log \left(\sum_{u}  \theta (x_1 , x_2, u)\right),
	\end{align}
	where the optimal parameters satisfy
	\begin{align}
		\label{3muequal1}
		& \sum_{x_1 , x_2} \mu_1 = \sum_{x_1 , x_2} \mu_2 =  \sum_{x_1 , x_2} \mu_0 = 1 .
	\end{align}
\end{lemma}

According to the previous discussion about Gaussian assumption in Section \ref{secIII},  we focus on the characterization of Gaussian random variables in this section. Let the source be a pair of jointly Gaussian random variables with zero mean and covariance matrix
\begin{align}\label{covx}
  C_{X_1,X_2} = \left[ {\begin{array}{*{10}{c}}
    1     & \rho\\
    \rho  & 1
    \end{array}} \right],
\end{align}
where $0 < \rho < 1 $. Moreover, the auxiliary random variables $ U, \hat X_1 , \hat X_2$ are all Gaussian distributed because of the linear Gaussian encoders and decoders. MSE is used to measure the distortion, let $d(x_1,\hat x_1) = (x_1 - \hat x_1 )^2$ and $d(x_2,\hat x_2) = (x_2 - \hat x_2 )^2$. Without loss of generality, we define $D_1^{max} \triangleq min_{\hat {x_1}} \mathbb{E} \{  d (x_1 , \hat {x_1})  \} = 1   $ and $D_2^{max} \triangleq min_{\hat {x_2}} \mathbb{E} \{ d (x_2 , \hat {x_2})  \} = 1   $, which is the same as Section IV of \cite{nayak2009successive}.

Before solving the dual optimal problem $ \mathcal{L}^{*} $, it is necessary to assume and define some useful variables for further derivation. When the parameters $( \boldsymbol{\alpha} , D_1 , D_2 ) \in {\cal D}_{\boldsymbol{\alpha},\boldsymbol{D}} ^{NZ}$, we firstly assume the relationship between the reconstructed sources and the auxiliary variable $U$, $V_1$ and $V_2$.
\newtheorem{assumption}{Assumption}
\begin{assumption}
According to the system model, the reconstruction $\hat X_1$ is the function of $U$ and $V_1$, $\hat X_2$ is symmetrical. Together with the assumption of Gaussian linear encoders and decoders, let
\begin{align}
   \label{hatx1}
   \hat X_1 & = m_1  U + V_1, \\
   \label{hatx2}
   \hat X_2 & = m_2  U + V_2,
   \end{align}
   where variables $ V_1$ and  $V_2$ are all Gaussian random variables with zero mean and independent of $U$.
\end{assumption}

Also due to the Gaussian linear encoders and decoders, the linear combination of sources $X_1, \ X_2$ and auxiliary variable $U$ is still Gaussian random variables. We make the following assumption
\begin{assumption}
Because the sources $X_1, \ X_2$ and auxiliary variable $U$ are all zero mean Gaussian random variables, we assume that $X_1 - m_1 U$ is Gaussian distribution follow $N(0,\frac{1}{(2 \omega_1)})$, while $X_2 - m_2 U$ is Gaussian distribution follow $N(0,\frac{1}{(2 \omega_2)})$. Where $\omega_1$ and $\omega_2$ are auxiliary real value.
\end{assumption}

At last, for simplicity, we define some parameters in next assumption.
\begin{assumption}
\begin{align}
& \widetilde \mu_i\triangleq \iint \mu_i \text{d} x_1  \text{d} x_2, \quad  \text{for} \  i = 0,1,2.  \\
\label{gamma1}
&    \eta_1  \triangleq \frac {b} {m_2^2} + 2 \frac{\alpha_2}{\alpha_0} \omega_2,
    \qquad
    \gamma_1 \triangleq \frac { \frac{\alpha_2}{\alpha_0} \omega_2 b} {m_1^2} + (\frac{\alpha_1}{\alpha_0} -1) \omega_1 \eta_1 , \\
    \label{gamma2}
&    \eta_2  \triangleq \frac {b} {m_1^2} + 2 \frac{\alpha_1}{\alpha_0} \omega_1 ,\qquad
    \gamma_2 \triangleq \frac { \frac{\alpha_1}{\alpha_0} \omega_1 b} {m_2^2} + (\frac{\alpha_2}{\alpha_0} -1) \omega_2 \eta_2,
\end{align}
where $b$ is an auxiliary positive real value. And in this section, we denote $\widetilde\theta (x_1, x_2) ,\ \widetilde\psi_1 (x_1, x_2,u) $, $ \widetilde\psi_2 (x_1, x_2,u) $ be continuous function which is similar as \eqref{tildetheta} just change sum symbol to integral symbol.
\end{assumption}

Utilizing the above assumptions, the dual optimal problem $ \mathcal{L}^{*} $ is solved in next lemma together with the alternative characterization of the dual problem in Lemma \ref{Lforalternative}.

\begin{lemma}\label{lemma_Ltransform}
The solution of $ \mathcal{L}^{*} $ is
\begin{align}\label{Ltransform}
\mathcal{L}^{*} = \alpha_0 \left(  h( \mathbf{X} ) + \log K_0 -   \frac{1}{2} \mathbb{E} \left\{ b ( \frac{X_1}{m_1} -  \frac{X_2}{m_2})^2 \right\}     \right)
\end{align}
where
\begin{align}\label{K0result}
&K_0  = \frac {1} {2 \pi} \sqrt{ |2 (\gamma_1 +  \eta_1  \omega_1)| ^ { 1- \frac{\alpha_1}{\alpha_0} - \frac{\alpha_2}{\alpha_0} } | 2 (\gamma_1 +  \eta_1 \frac{\beta_1}{\alpha_1})|^{\frac{\alpha_1}{\alpha_0}}  |2 (\gamma_2 +  \eta_2 \frac{\beta_2}{\alpha_2})|^{\frac{\alpha_2}{\alpha_0}} }. \\
& \mathbf{X} = [ X_1 \ \ X_2 ]^T.
\end{align}

\begin{IEEEproof}
The detailed proof is provided in Appendix \ref{prooflemmaK0}. Here we give the brief process of the proof. Firstly, we use the method of convolution to transform the parameters $\theta(x_1,  x_2, u)$, $\psi_1(x_1, x_2,u,\hat x_1)$ and $\psi_2(x_1, x_2,u,\hat x_2)$ to the form of probability density function. Then, utilizing the characterization of Gaussian random variables and the contractions in Lemma \ref{Lforalternative}, we can obtain $K_0$ by solving the equation sets.
\end{IEEEproof}
\end{lemma}

\begin{remark}\label{remark_M}
From the proof of above lemma in Appendix \ref{prooflemmaK0}, there are some interesting accompanying results. Matrix $\mathbf{H}^{-1} $ in \eqref{H} is the covariance matrix of $\mathbf{x} - [m_1 \ m_2]^T u$. While matrix $\mathbf{G_1}^{-1} $ in \eqref{G_1} is the covariance matrix of $\mathbf{x} - \mathbf{A_1} [u \ \hat x_1]^T $, where $\mathbf{A_1} $ is computed in \eqref{A_1}. Therefore we connect the dual optimal problem $\mathcal {L} ^*$ with Gaussian random variables, which is important and efficient to compute $\mathcal {L} ^*$ and RD function.
\end{remark}

\begin{remark}\label{condition=0}
    According the proof of above lemma in Appendix \ref{prooflemmaK0}, rewrite the solution of optimal $\mathcal{L} ^*$ in \eqref{solutionV1=0} and \eqref{solutionV2=0} here,
\begin{align*}
  & \gamma_1 ( \frac{\beta_1} {\alpha_1} - \omega_1 ) = 0  , \\
  & \gamma_2 ( \frac{\beta_2} {\alpha_2} - \omega_2 ) = 0  ,
\end{align*}
    This result is necessary for optimal $\mathcal{L} ^*$, which is equal to the constraint in \eqref{gamma1_the2} and \eqref{gamma2_the2}, and important for us to discuss all possible simulations for the final RD function.
\end{remark}

With the proof of above lemma and discussion in Remark \ref{remark_M}, we rewrite the covariance matrix of $(X_1, X_2)$ in next lemma.
\begin{lemma}
An alternative form of covariance matrix for the source $(X_1, X_2) $ is
\begin{align}
\label{cx1x2} &  \mathbf{C_{X_1,X_2}} =
   \left[ {\begin{array}{*{10}{c}}
    m_1^2 \sigma^2_U + \frac{\eta_1 }{2(\gamma_1 + \omega_1 \eta_1)}     & m_1 m_2 \sigma^2_U + \frac{ \frac{b}{m_1 m_2} }{2(\gamma_1 + \omega_1 \eta_1)} \\
    m_1 m_2 \sigma^2_U + \frac{ \frac{b}{m_1 m_2} }{2(\gamma_1 + \omega_1 \eta_1)}   & m_2^2 \sigma^2_U + \frac{\eta_2 }{2(\gamma_1 + \omega_1 \eta_1)}
    \end{array}} \right]  ,
\end{align}
\end{lemma}
\begin{IEEEproof}
 According to the expression of $ \mathbf{A_1} $ and $\mathbf{A_1}$ in proof of lemma \ref{lemma_Ltransform}, we can obtain that $\mathbf{A_1} [1 \ m_1]^T = \mathbf{A_2} [1 \ m_2]^T = [m_1 \ m_2]^T $. Together with the fact that \eqref{H} is the covariance matrix of $\mathbf{x} - [m_1 \ m_2]^T u$ in Remark \ref{remark_M}, the covariance matrix $\mathbf{C_{X_1,X_2}}$ is derived in another way with the variance of $U$ as $ \mathbf{C_{X_1,X_2}} = \mathbf{H^{-1}} + \sigma^2_U [m_1, m_2 ] [m_1, m_2 ]^T$,
which is equal to \eqref{cx1x2} by simply derivation and the lemma is proved.
\end{IEEEproof}

\begin{remark}
 With the fact that the matrix $\mathbf{G_1}^{-1} $ in \eqref{G_1} is the covariance matrix of $\mathbf{x} - \mathbf{A_1} [u \ \hat x_1]^T $ in Remark \ref{remark_M} and the assumption that $D_1 = \mathbb{E} (X_1 - \hat X_1)^2  $, the distortion $D_1$ is derived to
 \begin{align}
\label{D1} D_1 & = \sigma_{N_{11}}^2 =\frac {\eta_1}{2(\gamma_1 + \frac{\beta_1}{\alpha_1} \eta_1)}   .
\end{align}
 And
 the optimal $\beta_1$ is represented with auxiliary parameters via:
\begin{align}\label{optbeta}
 \beta_1   = \frac{\alpha_0 (1-m_1^2 \sigma^2_U)(m_1 -m_2 \rho) |\mathbf{H}|  }{2  m_1 D_1} .
\end{align}
The value of $D_2$ and $\beta_2$ is symmetrical.
\end{remark}

At last, together with \eqref{covx}, all the auxiliary parameters $b, \ \omega_i, \eta_i, \gamma_i \ i = 1, \ 2 $ can be replaced by three auxiliary parameters $ m_1, \ m_2 , \ \sigma^2_U $ and known parameters. Theorem 1 is proved by doing simply derivation of $\mathcal{L}^{*}$ in Lemma \ref{lemma_Ltransform}.

\section{Proof of Theorem \ref{finalRRR}}\label{Proof_th2}

According to the condition in Theorem \ref{simplyR}, we only need to determine the value of $m_1$, $m_2$ and $\sigma_U^2$ when $D_1$ and $D_2$ are given. We will discuss four regions of $ (D_1 , \ D_2) $, which the choice of $ ( m_1 , \ m_2 , \  \sigma_U^2) $ is different.
The four distortion regions are defined in \eqref{region_D1}-\eqref{region_D4}.

\subsection{ Determination of the Four Distortion Regions }\label{V.A}
In this section, we construct the distortion region with the constraint \eqref{gamma1_the2} and \eqref{gamma2_the2} in Theorem \ref{simplyR}. Referring to \eqref{omegabeta1}, the sign of $ ( 1- m_i^2 \sigma_U^2 - D_i )$ is the same as the variance of $V_i$, which is nonnegative. Next, according to \eqref{gamma1_the2} and \eqref{gamma2_the2}, we can divide the constraint into four parts as
\begin{align}
\label{condition_1} &f_{\alpha_1} (m_1, m_2) \ge 0 , \ f_{\alpha_2} (m_1, m_2) = 0, \ ( 1- m_1^2 \sigma_U^2 - D_1 ) = 0, \ ( 1- m_2^2 \sigma_U^2 - D_2 ) > 0. \\
\label{condition_2} &f_{\alpha_1} (m_1, m_2) = 0 , \ f_{\alpha_2} (m_1, m_2) \ge 0, \ ( 1- m_1^2 \sigma_U^2 - D_1 ) > 0, \ ( 1- m_2^2 \sigma_U^2 - D_2 ) = 0. \\
\label{condition_3} &f_{\alpha_1} (m_1, m_2) \ge 0 , \ f_{\alpha_2} (m_1, m_2) \ge 0, \ ( 1- m_1^2 \sigma_U^2 - D_1 ) = 0, \ ( 1- m_2^2 \sigma_U^2 - D_2 ) = 0. \\
\label{condition_4} &f_{\alpha_1} (m_1, m_2) = 0 , \ f_{\alpha_2} (m_1, m_2) = 0, \ ( 1- m_1^2 \sigma_U^2 - D_1 ) > 0, \ ( 1- m_2^2 \sigma_U^2 - D_2 ) > 0.
\end{align}

Because we focus on the general situation of $ (\boldsymbol{\alpha} , D_1 , D_2) \in {\cal D}_{\boldsymbol{\alpha},\boldsymbol{D}} ^{NZ} $, which is defined in Section \ref{secIII}, the distortion regions are established as following.
\begin{itemize}
\item For the situation \eqref{condition_1}, with the random variable assumption in \eqref{hatx1} and the variance of auxiliary random variable $V_1$ in \eqref{omegabeta1}, we obtain that $\sigma_{V_1} ^ 2 = 0 $ when $1- m_1^2 \sigma_U^2 - D_1 = 0$. The physical implication of this situation is that the rate on first-layer is zero, and messages are only communicated through second-layer and third-layer. Therefore letting $m_1 = 1$ will not change the value of RD function. Moreover, the value of $m_2^*$ is directed obtained from solving the equation  $f_{\alpha_2} (1, m_2) = 0$, where the existence and uniqueness is proved in Appendix \ref{appc}. At last, together with the constraint of $f_{\alpha_1} (m_1, m_2) \ge 0$, $( 1- m_2^2 \sigma_U^2 - D_2 ) > 0$, \eqref{con2}, the distortion region is established as \eqref{region_D1}.
\item For the situation \eqref{condition_2}, symmetric with the first distortion region \eqref{region_D1}, the second part of distortion region \eqref{region_D2} is easy to establish.
\item For the situation \eqref{condition_3}, because $ 1- m_1^2 \sigma_U^2 - D_1  = 0$ and $  1- m_2^2 \sigma_U^2 - D_2 = 0$, we determine $\sigma_U^2 =1 $ without loss of generality. Therefore,  the third distortion region is established by combining $m_i = \sqrt{ 1 - D_i }$, which is direct from above equations, with $f_{\alpha_1} (m_1, m_2) \ge 0$ and $ f_{\alpha_2} (m_1, m_2) = 0$.
\item For the situation \eqref{condition_4}, the parameters $m_1$ and $m_2$ are computable for solving the equation sets $f_{\alpha_1} (m_1, m_2) =  f_{\alpha_2} (m_1, m_2) = 0$. Therefore, according to $ D_1 < 1- m_1^2 \sigma_U^2$ and \eqref{con2}, the region of $D_1$ is determined as in \eqref{region_D4}. For $D_2$, it is symmetric.
\end{itemize}
\begin{remark}
The physical implication of the distortion region of \eqref{region_D1} is that the rate on first-layer is zero, and messages only communicate through second-layer and third-layer. In this situation, the Gray-Wyner system degenerates into a successive refinement system. Therefore, the distortion region is not only suitable for $ (\boldsymbol{\alpha} , D_1 , D_2) \in {\cal D}_{\boldsymbol{\alpha},\boldsymbol{D}} ^{NZ} $, but also for $\alpha_1 > \alpha_0$. Similarly, this fact is also applicable for \eqref{region_D2} because of symmetry.
\end{remark}

\subsection{ Solution of RD Function }
Next, we are ready to find pairs of set $(m_1, m_2, \sigma_U^2)$ which achieve RD function with constricts in Theorem \ref{finalRRR}.

\begin{itemize}
\item Case 1: $ (D_1 , \ D_2)  \in D_{ \boldsymbol{D}}^1   $:
In this distortion region, one of the choices of parameters set is
\begin{align}\label{choose1}
(m_1 , m_2 , \sigma_U^2) = ( 1, m_2^* , 1-D_1  ).
\end{align}
Because this choice implies that $ 1 - m_1^2  \sigma_U^2 = D_1$, which satisfies \eqref{gamma1_the2} in Theorem \ref{simplyR}. Obviously, the choice of $m_2 = m_2^*$ together with $ 1 - m_1^2  \sigma_U^2 = D_1$ makes sure that $ f_{\alpha_2} (1 , m_2^*) = 0  $ which satisfies \eqref{gamma2_the2}. And the condition \eqref{con2} implies that $D_1 \geq 1 -  { \rho} / { m_2^* }$, which is true with $D_1 > \frac { \frac{\alpha_1}{\alpha_0} ( m_{2}^* - \rho )^2 } { \frac{\alpha_1}{\alpha_0} ( m_{2}^{*2} - 2 m_{2}^* \rho + 1 ) + m_{2}^* \rho - 1 } $.
Therefore, RD function \eqref{Rcom} is achieved when choosing \eqref{choose1}.

\item Case 2: $ (D_1 , \ D_2)   \in D_{ \boldsymbol{D}}^2  $:
Using the same method as Case 1, the choice of parameters set
\begin{align}\label{choose2}
(m_1 , m_2 , \sigma_U^2) = ( m_1^*, 1 , 1-D_2  )
\end{align}
can obtain RD function of \eqref{Rcom} and satisfy \eqref{gamma1_the2}-\eqref{con2} in this distortion region.

\item Case 3: $ (D_1 , \ D_2)  \in D_{ \boldsymbol{D}}^3 $:
  Under this distortion region, we choose the parameters set as
  \begin{align}\label{choose3}
(m_1 , m_2 , \sigma_U^2) = ( \sqrt{1-D_1} , \sqrt{1-D_2} , 1  ).
\end{align}
The distortion region implies that
\begin{align}
\label{d1d2lep}& (1 - D_1 ) (1 - D_2)  \leq \rho^2, \\
\label{mind1d2gep} &\min  \bigg\{    \frac { 1 - D_1 } { 1 - D_2 }  , \frac { 1 - D_2 } { 1 - D_1 }  \bigg\} > \rho^2.
\end{align}
Because when ${ (1 - D_1 )} /{ (1 - D_2) } \leq \rho^2 $ yields that $ { (1 - D_2) }/ {( 1 - D_1 )} > {1}/{\rho^2} >  \rho^2 $ since $0<\rho<1$. While with the fact that $ 0 <  {\alpha_1} /{\alpha_0} \le 1 $, this situation is contradiction with $ f_{\alpha_1} (\sqrt{1-D_1} , \sqrt{1-D_2}) > 0$. Therefore, \eqref{d1d2lep} and \eqref{mind1d2gep} implies \eqref{con2}. \eqref{gamma1_the2} and \eqref{gamma2_the2} are also obvious to obtain after simple computation as $1 - m_1^2 \sigma^2_U - D_1 = 1 - m_2^2 \sigma^2_U - D_2 = 0$. Therefore, the choice of \eqref{choose3} can achieve the     RD function in \eqref{Rcom}.

\item Case 4: $ (D_1 , \ D_2)  \in D_{ \boldsymbol{D}}^4 $ \\
\textbf{Subcase A.} $a_1 \neq a_2 $: Under this distortion region, we can choose parameters set as
\begin{align}
 (m_1, m_2 , \sigma_U^2) =  ( \sqrt{1-\nu_1} , \sqrt{1-\nu_2} , 1  )
\end{align}
where $\nu_1$ and $\nu_2$ are defined in \eqref{nu1} and \eqref{nu2}. With this value, we obtain that
\begin{align*}
  (\rho - m_1 m_2 \sigma^2_U)^2 = \frac {( \alpha_0 - \alpha_1 )( \alpha_0 - \alpha_2 )} {\alpha_1 \alpha_2} (1 - m^2_1 \sigma^2_U) (1 - m^2_2 \sigma^2_U)   , \\
  \frac { m_2 } { m_1 } = \frac { (\alpha_1 - \alpha_2) \rho + \sqrt { (\alpha_1 - \alpha_2)^2 \rho^2 + 4 ( \alpha_0 - \alpha_1 )( \alpha_0 - \alpha_2 )  } } {2 ( \alpha_0 - \alpha_2 ) }  ,  \\
  \frac { m_1 } { m_2 } = \frac { (\alpha_2 - \alpha_1) \rho + \sqrt { (\alpha_1 - \alpha_2)^2 \rho^2 + 4 ( \alpha_0 - \alpha_1 )( \alpha_0 - \alpha_2 )  } } {2 ( \alpha_0 - \alpha_1 ) }  .
\end{align*}
With these results putting into the function of $f_{\alpha_1}(m_1, m_2)$ and $f_{\alpha_2}(m_1, m_2)$ defined in \eqref{f1_the2} and \eqref{f2_the2}, $f_{\alpha_1}(m_1, m_2) = f_{\alpha_2}(m_1, m_2) =0 $ is obtained, implying \eqref{gamma1_the2} and \eqref{gamma2_the2}. Moreover, \eqref{con2} is satisfied after the straightforward computation of plugging $m_1$ an $m_2$ in \eqref{condition_4}.
Therefore, the RD function is achievable and we propose a simple expression here
\begin{align*}
  R_{\boldsymbol{\alpha}}(D_1,D_2)
& = \frac{\alpha_0}{2} \log \frac{1-\rho^2}{ \frac { \alpha_0 \alpha_1 + \alpha_0 \alpha_2 - \alpha^2_0  } { \alpha_1 \alpha_2 } \nu_1 \nu_2}  + \frac{\alpha_1}{2} \log \frac{\nu_1 }{D_1} + \frac{\alpha_2}{2} \log \frac{\nu_2}{D_2}.
\end{align*}

\textbf{Subcase B}. $\alpha_1 = \alpha_2 $:
Under this distortion region, we can choose parameters set as
\begin{align}
(m_1, m_2 ,\sigma_U^2)  =  (\sqrt{\left| \frac {\alpha_1 \rho + \alpha_1 - \alpha_0} {\alpha_0 - \alpha_1 - \alpha_2 } \right| } , \sqrt{\left| \frac {\alpha_2 \rho + \alpha_2 - \alpha_0} {\alpha_0 - \alpha_1 - \alpha_2 }\right| } , 1)
\end{align}

With this value, it is given that
\begin{align}
    & 1 - m_1^2 \sigma_U^2 = \frac{\alpha_1 (1-\rho)}{\alpha_1+\alpha_2 - \alpha_0}  ,  \\
    & 1 - m_2^2 \sigma_U^2 = \frac{\alpha_2 (1-\rho)}{\alpha_1+\alpha_2 - \alpha_0}  ,  \\
  \label{p=m1m20}&  \rho - m_1 m_2 \sigma_U^2 = \frac{(1-\rho)(\alpha_0-\alpha_1)}{\alpha_1+\alpha_2 - \alpha_0}.
\end{align}
With these results putting into the function of $f_{\alpha_1}(m_1, m_2)$ and $f_{\alpha_2}(m_1, m_2)$ defined in \eqref{f1_the2} and \eqref{f2_the2}, $f_{\alpha_1}(m_1, m_2) = f_{\alpha_2}(m_1, m_2) =0 $ is obtained, implying \eqref{gamma1_the2} and \eqref{gamma2_the2}. Moreover, \eqref{con2} is satisfied after the straightforward computation of plugging $m_1$ an $m_2$ in \eqref{condition_4}.
\end{itemize}

\section{Conclusion}
In this paper, the main contribution of this paper is proposing an analytical expression of the RD function for Gray-Wyner lossy source coding system with Gaussian sources under the constraint of quadratic distortion. We transform the problem of computing the RD function into a dual optimal problem to obtain some results and conditions for the optimal problem. Furthermore, we introduce some useful intermediate variables to establish the connection between the random variables and the optimal probability density function. Based on the main results and proofs, we provide an alternative method to compute Wyner's CI through the proposed RD function. Finally, we provide numerical simulations of the proposed iterative algorithm to show the convergence and RD function for different parameters on a joint Gaussian source, for making the expression intuitive.

\subsection{Proof of Lemma \ref{Lforalternative}}\label{Proof_l4}

In this section, we come to prove the lower bound of $ \mathcal{L}^* $.  Tt remains to prove the converse part of this optimal problem as
    \begin{align}\label{uallcompute}
    \mathcal{L}^{P} \geq - \alpha_0 \sum_{x_1 , x_2} p(x_1 , x_2) \log \left(\sum_{u}  \theta (x_1 , x_2, u)\right),
    \end{align}
for any $ p(u , \hat x_1, \hat x_2 | x_1 , x_2 ) $ and $ ( \theta(x_1 , x_2, u) , \psi_1(x_1 x_2,u,\hat x_1) , \psi_2(x_1, x_2,u,\hat x_2) ) \in \Gamma$.

Furthermore, we aim to confirm the truth of \eqref{uallcompute}. For simplicity, we define
\begin{align}\label{tildetheta}
&\widetilde \theta(x_1 , x_2) \triangleq \sum_{u}  \theta (x_1 , x_2, u), \nonumber\\
&\widetilde \psi_1(x_1 , x_2, u) \triangleq \sum_{\hat x_1}  \psi_1 (x_1 , x_2, u , \hat x_1), \\
&\widetilde \psi_2(x_1 , x_2, u) \triangleq \sum_{\hat x_2}  \psi_2 (x_1 , x_2, u , \hat x_2).  \nonumber
\end{align}
 then \eqref{uallcompute} is derived via
  \begin{align}\label{u1compute}
  & \mathcal{L}^{P} + \alpha_0 \sum_{x_1 , x_2} p( x_1 , x_2 ) \log \widetilde \theta(x_1 , x_2)     \nonumber \\
    & \quad \mathop \ge \limits ^ {(a)}   \sum_{x_1 , x_2, u , \hat x_1 , \hat x_2} p ( x_1 , x_2 , u , \hat x_1 , \hat x_2  ) \left[  \alpha_0  ( 1 - \frac {\mu_0}{ p(x_1, x_2 |u) } )
  + \alpha_1 \bigg ( 1 - \frac { p ( \hat x_1   | u )  \mu_1 } { p ( \hat x_1   | x_1,x_2 , u ) \mu_0 } \bigg)
   + \alpha_2 \bigg ( 1 - \frac { p ( \hat x_2   | u )  \mu_2 } { p (  \hat x_2  | x_1 , x_2 ,u ) \mu_0 } \bigg) \right] \nonumber\\
  & \quad  = \alpha_0 \left( 1 - \sum_{ u } p(u) \left(\sum_{ x_1, x_2 } \mu_0 \right) \right)
  + \sum_{i=1}^{2} \alpha_i  p ( x_1 , x_2 , u,  \hat x_1 , \hat x_2 ) \sum_{x_1 , x_2, u , \hat x_1 } \frac{p(x_1 , x_2 | u)}{\mu_0} \left( \frac{\mu_0 } {p(x_1 , x_2 | u)} - \frac{\mu_i } {p(x_1 , x_2 | u, \hat x_i)}  \right) \nonumber \\
  \end{align}
  The inequality $(a)$ follows from the inequality $  \log \tau \geq 1 - 1/{\tau}  $, and holds if and only if $\tau = 1$.
  Since $ \sum_{x_1, x_2 } \mu_0 \leq 1$, the first part of \eqref{u1compute} is bounded
  \begin{align}\label{u0==1}
   \alpha_0 \left( 1 - \sum_{ u } p(u) \left(\sum_{ x_1, x_2 } u_0 \right) \right) \geq 0
  \end{align}
  with equality when $\sum_{x_1, x_2 } \mu_0 = 1$. Then we consider the second part of \eqref{u1compute} using the fact ${p(x_1 , x_2 | u)}/{\mu_0} > 0  $
  \begin{align}\label{u0u1u2=1}
  &\alpha_1 p ( x_1 , x_2 , u,  \hat x_1 , \hat x_2 )  \sum_{x_1 , x_2, u , \hat x_1 } \frac{p(x_1 , x_2 | u)}{\mu_0} \left( \frac{\mu_0 } {p(x_1 , x_2 | u)} - \frac{\mu_1 } {p(x_1 , x_2 | u, \hat x_1)}  \right)  \nonumber\\
  & \quad \mathop \ge \limits ^ {(b)} \alpha_1   \left(  \sum_{x_1 , x_2, u , \hat x_1 } \xi \right)  \left(  \sum_{ u } p(u) \left(\sum_{ x_1, x_2 } \mu_0 \right)  -  \sum_{ u , \hat x_1 } p(u, \hat x_1) \left(\sum_{ x_1, x_2 } \mu_1 \right)   \right) \nonumber\\
  & \quad \mathop \ge \limits ^ {(c)} 0,
  \end{align}
  where $\xi$ is denoted as the minimum of $ {p(x_1 , x_2 | u)}/{\mu_0} $. Therefore, inequality (b) is true, and the correctness of inequality (c) is from the fact that $\sum_{x_1 , x_2} \mu_1 \leq \sum_{x_1 , x_2} \mu_0 \leq 1$. Therefore, this inequality holds if $\sum_{x_1 , x_2} \mu_i = \sum_{x_1 , x_2} \mu_0$. Now, we can claim that the inequality \eqref{uallcompute} is true and with equality if $\sum_{x_1 , x_2} \mu_1 = \sum_{x_1 , x_2} \mu_2 =  \sum_{x_1 , x_2} \mu_0 = 1$. So, we obtain $\mathcal{L}^{P} + \alpha_0 \sum_{x_1 , x_2} p( x_1 , x_2 ) \log \widetilde \theta(x_1 , x_2) \geq 0 $ and prove Lemma \ref{Lforalternative}.

\section{Proof of Lemma \ref{lemma_Ltransform} }\label{prooflemmaK0}
Firstly, the formula \eqref{Ltransform} is obtained directly by expanded parameters in \eqref{Lfunctionoftheta}, where the original $K_0$ is the solution of
\begin{align}\label{assumeK0}
 &\frac { p ( x_1 , x_2 ) } { \widetilde \theta ( x_1 , x_2p ) }= K_0 e^{ - \frac{1}{2}  b ( \frac{x_1}{m_1} -  \frac{x_2}{m_2})^2 }.
\end{align}

To solve $K_0$ in \eqref{Ltransform}, we derive $\widetilde \psi_i (x_1, x_2 , u)$ using the method of convolution together with \eqref{hatx1} with Assumption 2, via
\begin{align}
 \label{psi1} &\widetilde \psi_i ( x_1 , x_2 , u ) = {\sqrt { \frac { \omega_i } { \frac {\beta _i} {\alpha_i} } } } e ^{- \omega_i { ( x_i - m_i u) } ^ 2 },
\end{align}
where the variances of $V_1$ and $V_2$ defined in \eqref{hatx1} are determined as
\begin{align}
  \label{omegabeta1} 2  \sigma_{V_i}^2 = \frac{1} { \omega_i } - \frac{\alpha_i}{\beta_i} \ge 0.
\end{align}

 Then we derive the continuous form of $ \mu_0 (u) $, $\mu_1 ( u , \hat x_1 )$ and $\mu_2 ( u , \hat x_2 )$, 
 with the assumption of $K_0$, $\widetilde \psi_1(x_1, x_2, u)$ and $\widetilde \psi_2(x_1, x_2, u)$ in \eqref{assumeK0}-\eqref{psi1} respectively:
 \begin{small}
\begin{align}
  \label{tildemu0}
  \widetilde \mu_0  & = K_0  { \sqrt {\frac { \omega_1 } { \frac {\beta _1} {\alpha_1} } } } ^ { \frac {\alpha _1} {\alpha_0} } { \sqrt {\frac {\omega_2} { \frac {\beta _2} {\alpha_2} } } } ^ { \frac {\alpha _2} {\alpha_0} } { \frac { 2 \pi } { \sqrt { | \mathbf{H} | } } } \cdot {\iint} { \frac { \sqrt { |\mathbf{H}| } }  {2 \pi} } e ^{ - \frac {1} {2} (\mathbf { x - s }  u )^T  \mathbf{H}  (\mathbf{ x - s } u ) } d x_1 d x_2   ,   \\
  \label{tildemu1}
  \widetilde \mu_1   &= K_0  { \sqrt { \frac { \omega_1 } { \frac {\beta_1} {\alpha_1} } } } ^ { ( \frac {\alpha_1} {\alpha_0} -1 ) } { \sqrt { \frac { \omega_2 } { \frac {\beta _2} {\alpha_2} } } } ^ {\frac {\alpha_2} {\alpha_0} } { \frac{2 \pi} { \sqrt { | \mathbf{G_1} | } } } e^ { - \frac {1} {2} \mathbf { \hat { z_1 } } ^ T \mathbf { F_1 \hat z_1 }  }      \cdot   {\iint} { \frac {\sqrt { |\mathbf {G_1} | } } {2 \pi} } e ^ { - \frac {1} {2} \mathbf { (x - A_1 \hat z_1)} ^T \mathbf { G_1 ( x - A_1 \hat z_1) }  }  d x_1 d x_2   ,  \\
  \label{tildemu2}
   \widetilde \mu_2 & = K_0  { \sqrt { \frac { \omega_1 } { \frac {\beta_1} {\alpha_1} } } } ^ {  \frac {\alpha_1} {\alpha_0}  } { \sqrt { \frac { \omega_2 } { \frac {\beta _2} {\alpha_2} } } } ^ { ( \frac {\alpha_2} {\alpha_0} - 1) } { \frac{2 \pi} { \sqrt { | \mathbf{G_2} | } } } e^ { - \frac {1} {2} \mathbf { \hat { z_2 } } ^ T \mathbf { F_2 \hat z_2 }  }      \cdot   {\iint} { \frac {\sqrt { |\mathbf {G_2} | } } {2 \pi} } e ^ { - \frac {1} {2} \mathbf { (x - A_2 \hat z_2)} ^T \mathbf { G_2 ( x - A_2 \hat z_2) }  }  d x_1 d x_2 ,
\end{align}
\end{small}
where
\begin{align}
	\label{H} & \mathbf{H}  = \left[ {\begin{array}{ccc}
			\frac {b} { m_1^2 }  +  2 \frac{ \alpha_1 } { \alpha_0 } \omega_1 &   -\frac{b}{m_1 m_2}  , \\
			-\frac {b} { m_1 m_2}   &    \frac{b}{m_2^2} +  2\frac{\alpha_2}{\alpha_0} \omega_2
	\end{array}} \right]   ,  \\
	\label{G_1}  &\mathbf{G_1} =
	\mathbf{H} +
	[ 2 \frac{\beta_1}{\alpha_1}  -2 \omega_1  \ \ 0]^T [ 1 \ \ 0  ]
	,  \\
	\label{A_1} & \mathbf{A_1} =  \mathbf{G_1^{-1}}   \left[ {\begin{array}{ccc}
			2 \omega_1 (\frac{\alpha_1}{\alpha_0} -1) m_1   &  2 \frac{\beta_1}{\alpha_1} \\
			2 \omega_2 \frac{\alpha_2}{\alpha_0} m_2        &  0
	\end{array}} \right]  ,  \\
	\label{F_1} & \mathbf{F_1} =
	\frac{2 \frac{\beta_1}{\alpha_1} \gamma_1}  {\gamma_1 + \eta_1 \frac{\beta_1}{\alpha_1}}
	\left[ {\begin{array}{ccc}
			m_1^2   &    -m_1   \\
			-m_1     &   1
	\end{array}} \right] ,
\end{align}
and $\mathbf{s}  = [m_1 \quad m_2]^T$, $\mathbf{\hat Z_1} = [U \quad \hat X_1]^T$. Moreover, for $\mathbf{G_2}$, $\mathbf{A_2}$ , $\mathbf{F_2}$ and $\mathbf{\hat Z_1}$, it is symmetric. With the fact that the solution of $\mathcal{L}^{*}$ implies $ \widetilde \mu_0 = \widetilde \mu_1 = \widetilde \mu_2 = 1$ in Lemma \ref{Lforalternative}, the function in the integral of \eqref{tildemu0} can be seen as the probability density function of Gaussian random vector $ \mathbf{x} - \mathbf{s} u  $. Similarly the function in the integral of \eqref{tildemu1} can be seen as the probability density function of Gaussian random vector $ \mathbf{x} - \mathbf{A_1} \hat z_1  $, and it is symmetric for \eqref{tildemu2}. Therefore the equation set $ \widetilde \mu_0 = \widetilde \mu_1 = \widetilde \mu_2 = 1$ can determine $K_0$ letting the coefficient, before integral symbol, be $1$ in \eqref{tildemu0}, \eqref{tildemu1} and \eqref{tildemu2}.

One of the solutions of this equation set is
\begin{align}\label{solutionV1=0}
  & \gamma_1 ( \frac{\beta_1} {\alpha_1} - \omega_1 ) = 0  , \\
  \label{solutionV2=0}
  & \gamma_2 ( \frac{\beta_2} {\alpha_2} - \omega_2 ) = 0  ,
\end{align}
which can make sure that
$e^ { - \frac {1} {2} \mathbf { \hat { z_1 } } ^ T \mathbf { F_1 \hat z_1 }  } = e^ { - \frac {1} {2} \mathbf { \hat { z_2 } } ^ T \mathbf { F_2 \hat z_2 }  } =1$. Therefore, 
the solution of $K_0$ is
\begin{align}\label{solutioncor3}
  K_0  = \frac {1} {2 \pi} \sqrt{|\mathbf{H}| ^ { 1- \frac{\alpha_1}{\alpha_0} - \frac{\alpha_2}{\alpha_0} } |\mathbf{G_1}|^{\frac{\alpha_1}{\alpha_0}}  |\mathbf{G_2}|^{\frac{\alpha_2}{\alpha_0}} }.
\end{align}
With the fact that $ |\mathbf{ H }| = 2 (\gamma_1 +  \eta_1  \omega_1) = 2 (\gamma_2 +  \eta_2  \omega_2) >0 $ , $ |\mathbf{ G_1 }| = 2 (\gamma_1 +  \eta_1 {\beta_1}/{\alpha_1})  >0 $ and $ |\mathbf{ G_2 }| = 2 (\gamma_2 +  \eta_2 {\beta_2}/{\alpha_2})  >0 $, $K_0$ in \eqref{solutioncor3} is equal to \eqref{K0result}. Therefore, this Lemma is proved.

\section{ The Unique Root on $f_{\alpha_2} (m_1, m_2) = 0 $  }\label{appc}
In this section, we prove that $f_{\alpha_2} (m_1, m_2) = 0 $ has only one root when $m_1 = \sqrt {\frac{1-D_1}{\sigma^2_U}}$.
When all parameters satisfy Theorem \ref{simplyR}, we derive the function $ f_{\alpha_2} (\sqrt {\frac{1-D_1}{\sigma^2_U} } , m_2) $ via
\begin{small}
\begin{align}
\label{rootf1}&f_{\alpha_2}  ( \sqrt {\frac{1-D_1}{\sigma^2_U} } ,  {\sqrt {\frac{1-D_1}{\sigma^2_U \rho^2} }}  ) = (\frac{\alpha_2}{\alpha_0} -1) \sqrt{\frac{1-D_1}{\sigma^2_U} } (  \frac{1}{\rho} - \rho)  (1- \frac{1-D_1}{\rho^2} ) , \\
\label{rootf2}&f_{\alpha_2}  (\sqrt {\frac{1-D_1}{\sigma^2_U} } , \frac {\rho} {\sqrt {  (1-D_1) \sigma^2_U} }  ) = (\frac{\alpha_2}{\alpha_0} -1) \rho \frac{1}{\sqrt {\sigma^2_U}} ( \sqrt {\frac{1}{1-D_1}} -  \sqrt{1-D_1} )  (1- \frac {\rho^2} {1-D_1} ),  \\
\label{rootf3}&f_{\alpha_2}  ( \sqrt {\frac{1-D_1}{\sigma^2_U} } , \sqrt {\frac{1-D_1}{\sigma^2_U} } \rho ) = \frac{\alpha_2}{\alpha_0} \sqrt {\frac{1-D_1}{\sigma^2_U} } \rho (1- \rho^2) D_1 .
\end{align}
\end{small}
Because the two constraints $m_1 \rho < m_2 < {m_1}/{\rho}$ and $m_1 \rho < m_2 < {\rho}/{(m_1 \sigma^2_U)} $ must be satisfied, we consider following three cases:\\
\begin{enumerate}
 \item \textit{$D_1 > 1- \rho^2$}: At this condition, it is easy to get that ${m_1}/{\rho} < {\rho}/{m_1 \sigma^2_U}$. Therefore $\eqref{rootf3} > 0 $, \eqref{rootf1}  has the same sign with ${\alpha_2}/{\alpha_0} -1 < 0$ and \eqref{rootf2} has the same sign with $ 1- {\alpha_2}/{\alpha_0} > 0 $. So there are only one root of $ m_2$ between $ m_1 \rho$ and $\frac{ m_1}{\rho} $.
 \item \textit{$D_1 < 1- \rho^2$}: As the same of above, we can get that ${\rho}/{ (m_1 \sigma^2_U) } < { m_1}/{\rho} $. Therefore $\eqref{rootf3} > 0 $, \eqref{rootf2}  has the same sign with ${\alpha_2}/{\alpha_0} - 1 < 0 $ and \eqref{rootf1} has the same sign with $ 1- {\alpha_2}/{\alpha_0} > 0 $. So there are only one root of $ m_2$ between $ m_1 \rho$ and $ {\rho}/{ (m_1 \sigma^2_U)}$.
 \item \textit{$D_1 = 1- \rho^2$}: As the same of above, we can get that ${\rho}/{ m_1 \sigma^2_U} = { m_1}/{\rho} $. Therefore $\eqref{rootf1} > 0 $ and $\eqref{rootf2}  =  \eqref{rootf3} = 0$. However, when $ m_2 = {1}/{\sqrt{\sigma^2_U}}$, it may cause $|\mathbf{H}| = 0$, which has contradiction with the results above. So we consider the function $h(m_2) =  {f_{\alpha_2} ( m_2 )}/ { (m_1 -  m_2 \rho)} $ which becomes a quadratic equation. It is obvious that $h( m_1 \rho)  < 0 $ and $h(  { m_1}/ {\rho} )  > 0 $, so there is only one root between $ m_1 \rho$ and $ ={\rho}/{( m_1 \sigma^2_U)}$.
\end{enumerate}
\ifCLASSOPTIONcaptionsoff
  \newpage
\fi
\bibliographystyle{IEEEtranTCOM}
\bibliography{IEEEabrv,../bib/paper}

\end{document}